\documentclass[11pt,fleqn]{article} 
\usepackage{psfig}
\title{Possible Frictionless States at Room-Temperature Regime \\ for  Many Fermions in Confined Domain}  %
\author{Zotin K.-H. Chu }
\date{
P.O. Box 39, Toudiban, Road Xihong, Urumqi 830000,  China} 
\begin{document}           
\maketitle
\doublerulesep=6.5mm        
\baselineskip=6.5mm
\oddsidemargin-1mm         
%
\begin{abstract}
We investigate the possible frictionless transport of many
composite (condensed) fermions at room temperature regime along an
annular tube with transversely wavy-corrugations by using the
verified transition-rate  model and boundary perturbation
approach. We found that for certain   activation volume and energy
there exist possible frictionless states at room temperature
regime.
\newline

\end{abstract}
%
\bibliographystyle{plain}
\section{Introduction}
Understanding the frictionless transport of system of (interacting
or noninteracting) fermions is crucial to our knowledge of the
nature considering the neutron star, the superconductivity, the
superfluidity, etc. One relevant study is the existence and
characteristics of the Fermi surface for a system of interacting
fermions. Usually such a surface is introduced into momentum space
only for a system of noninteracting fermions.  \newline To be
specific, this surface represents the limit of occupation of the
different single-particle momentum states in the ground state of
the system. All the states with momentum contained  within this
surface are occupied, all those with momentum outside this surface
are unoccupied. However, Migdal [1] noticed that, under some
circumstances, the mean occupation number of different
single-particle momentum states in the true ground state still
possessed a discontinuity for a system of interacting fermions.
Meanwhile, as Luttinger pointed out, the existence of the Fermi
surface depends on the nature of forces between the fermions [2].
\newline
Other relevant study is, e.g., thermodynamical properties of
trapped noninteracting Fermi gases in gravitational fields [3]
which was originated in view of the successful experiments with
the trapping and cooling of weakly interacting fermionic isotopes
[4]. There is no doubt that it is important to theoretically
investigate the thermodynamical and statistical properties of
system of fermions even there are many approaches [5-11]. Here we
shall investigate the possible frictionless transport of many
fermions in a confined domain via a different approach considering
the shearing response of many-fermion system. Note that
researchers have been interested in the question of how matter
responds to an external mechanical load. External loads cause
transport, in Newtonian or various types of non-Newtonian ways.
Amorphous matter, composed of polymers, metals, or ceramics, can
deform under mechanical loads, and the nature of the response to
loads often dictates the choice of matter in various applications.
To the best knowledge of the author, the simplest model that makes
a prediction for the rate and temperature dependence of shear
yielding is the rate-state model of stress-biased thermal
activation [12-14]. Structural rearrangement is associated with a
single energy barrier $E$ that is lowered or raised linearly by an
applied stress $\sigma$ : $R_{\pm}=\nu_0 \exp[-E/(k_B T)] \exp[\pm
\sigma V^*/(k_B T)],$ where $k_B$ is the Boltzmann constant,
$\nu_0$ is an attempt frequency and $V^*$ is a constant called the
'activation volume'. In amorphous  matter, the transition rates
are negligible at zero stress. Thus, at finite stress one needs to
consider only the rate $R_{+}$ of transitions in the direction
aided by stress.\newline The linear dependence will always
correctly describe small changes in the barrier height, since it
is simply the first term in the Taylor expansion of the barrier
height as a function of load. It is thus appropriate when the
barrier height changes only slightly before the system escapes the
local energy minimum. This situation occurs at higher
temperatures; for example, Newtonian transport is obtained in the
rate-state model in the limit where the system experiences only
small changes in the barrier height before thermally escaping the
energy minimum. As the temperature decreases, larger changes in
the barrier height occur before the system escapes the energy
minimum (giving rise to, for example, non-Newtonian transport). In
this regime, the linear dependence is not necessarily appropriate,
and can lead to inaccurate modeling.  To be precise, at low shear
rates ($\dot{\gamma} \le \dot{\gamma}_c$), the system behaves as a
power law shear-thinning material while, at high shear rates, the
stress varies affinely with the shear rate. These two regimes
correspond to two stable branches of stationary states, for which
data obtained by imposing either $\sigma$ or $\dot{\gamma}$
exactly superpose.
\newline
In this  paper,  considering the general forcing (either gravity
[3] or electric field [6]) we shall adopt the verified
transition-rate-state model [12-14] to study the frictionless
transport of many fermions within a corrugated annular tube.  To
obtain the law of shear-thinning matter for explaining the too
rapid annealing at the earliest time, because the relaxation at
the beginning was steeper than could be explained by the
bimolecular law, a hyperbolic sine law between the shear (strain)
rate : $\dot{\gamma}$ and  shear stress : $\tau$ was proposed and
the close agreement with experimental data was obtained. This
model has sound physical foundation from the thermal activation
process [12-14] (a kind of (quantum) tunneling which relates to
the matter rearranging by surmounting a potential energy barrier
was discussed therein). With this model we can associate the
(shear-thinning) fluid with the momentum transfer between
neighboring atomic clusters on the microscopic scale and reveals
the atomic interaction in the relaxation of flow with dissipation
(the momentum transfer depends on the activation (shear) volume :
$V^*\equiv V_h$ which is associated with the center distance
between atoms and is equal to $k_B T/\tau_0$ ($T$ is temperature
in Kelvin, and $\tau_0$ a constant with the dimension of stress).
\newline To consider the more realistic but complicated  boundary
conditions in the walls of the annular tube, however, we will use
the boundary perturbation technique [15-16] to handle the presumed
wavy-roughness along the walls of the annular tube. To obtain the
analytical and approximate solutions, here, the roughness is only
introduced in the radial or transverse direction. The relevant
boundary conditions along the wavy-rough surfaces will be
prescribed below.  We shall describe our approach after this
section : Introduction with the focus upon the transition-rate
approach and boundary perturbation method. The approximate
expression of the transport is then demonstrated at the end.
Finally, we will illustrate our results  and give discussions
therein.
\section{Formulations} %
We firstly take into account the condensed system of many fermions
subjected to random thermal fluctuations (under external forcing).
In thermally-activated motion, mobile fermions in a many-fermion
system may interact with other fermions even they are already in a
preferred motion. The rate of deformation (strain) is controlled
by the rate at which thermal energy can help the composite systems
overcome their energy barriers, allowing the rest of other
fermions to spread. \newline
 The attractive interactions involving pairs
of moving fermions lower the energy as the mechanisms that enable
them to form a composite moving subsystem cost energy. Repulsive
interactions, on the other hand, require less work to overcome,
and do not usually transform the composite subsystem or leave
residual subsystem after the interaction. Parts of condensed
fermions easily pull away after having been forced to pass the
repulsive obstacle by an external stress. Accordingly, attractive
interactions require thermal activation and are said to be
temperature-dependent, while repulsive ones are not.
\newline
In fact, thermal energy is supplied by random thermal
fluctuations, and  motion of composite (condensed) fermions
depends on the number of fluctuations that supply the interacting
subsystems the energy they need. The number of such successful
outcomes is $N_s = N P(\mbox{success})$, where $N$ is the number
of attempts in the complete many-fermion system. The probability
of success is the probability that the thermal jump $U_j$ is
greater than $\Delta E$, the energy required to surmount the
barrier and assumed to follow an Arrhenius law [12-14] which is
$P(\mbox{success})=P(U_j > \Delta E)=\exp(-\Delta E/k_B T)$.
Considering the strain gained at each successful attempt leads to
the thermal-activation controlled expression for the strain rate
$\dot{\gamma}_p=\dot{\gamma}_{p_0} \exp(-\Delta E(\tau_s)/k_B T)$.
Here $\dot{\gamma}_{p_0}$ is a intrinsic value that has units of
strain rate and depends on the average waiting time at the
intersection point, the strain released after the events, and the
frequency of thermal fluctuations, which is some fraction of the
Debye frequency ($10^{13}$ s$^{-1}$). We remind the readers that
$\Delta E(\tau_s)$ is a function of $\tau_s$ which is a
concentrated shear stress due to the short-range interaction with
another subsystem (or obstacle), emphasizing the localized nature
of thermally-activated events. It is possible to attribute $\Delta
E$ in above expression to one type of thermally-activated process.
This simplification applies when one process has a much smaller
$\Delta E$ than the rest. Thermally controlled  deformation,
however, is a complex collective phenomenon of many
thermally-activated processes. To continue to use this theory,
$\Delta E$ must be treated as an effective energy covering all
possible types of such processes.
\newline
We shall consider a steady transport of  many fermions in a
wavy-rough annular tube of $r_1$ (mean-averaged inner radius) with
the inner interface being a fixed wavy-rough surface :
$r=r_1+\epsilon \sin(k \theta+\beta)$ and $r_2$ (mean-averaged
outer radius) with the outer interface being a fixed wavy-rough
surface : $r=r_2+\epsilon \sin(k \theta)$, where $\epsilon$ is the
amplitude of the (wavy) roughness, $\beta$ is the phase shift
between two walls, and the roughness wave number : $k=2\pi /L $
($L$ is the wavelength of the surface modulation in
transverse direction). 
\newline Firstly, this  matter (composed of  many condensed (composite) fermions)  can be expressed
as [12-14]
 $\dot{\gamma}=\dot{\gamma}_0  \sinh(\tau/\tau_0)$,
where $\dot{\gamma}$ is the shear rate, $\tau$ is the shear
stress, $\tau_0=2 k_B T/V_h$, and $\dot{\gamma}_0 (\equiv C_k  k_B
T \exp(-\Delta E/k_B T)/h$) is with the dimension of the shear
rate; here $C_k \equiv 2 V_h/V_m$ is a constant relating rate of
strain to the jump frequency ($V_h=\lambda_2\lambda_3\lambda$,
$V_m=\lambda_2\lambda_3\lambda_1$, $\lambda_2 \lambda_3$ is the
cross-section of the transport unit on which the shear stress
acts, $\lambda$ is the distance  jumped  on each relaxation,
$\lambda_1$ is the perpendicular distance between two neighboring
layers of particles sliding past each other), accounting for the
interchain co-operation required, $h$ is the Planck constant,
$\Delta E$ is the activation energy. \newline In fact, the force
balance gives the shear stress at a radius $r$ as $\tau=-(r
\,\delta{\cal G})/2$ [15]. $\delta{\cal G}$ is the net effective
external (gravity or electric field) forcing along the transport
(or tube-axis : $z$-axis) direction (considering $dz$ element).
Introducing the forcing parameter
$\phi = -(r_2/2\tau_0) \delta{\cal G}$
then we have
 $\dot{\gamma}= \dot{\gamma}_0  \sinh ({\phi r}/{r_2})$.
As $\dot{\gamma}=- du/dr$ ($u$ is the velocity of the transport in
the longitudinal ($z$-)direction of the annular (cosmic) string),
after integration, we obtain
\begin{equation}
 u=u_s +\frac{\dot{\gamma}_0 r_2}{\phi} [\cosh \phi - \cosh (\frac{\phi r}{r_2})],
\end{equation}
here, $u_s (\equiv u_{slip})$ is the velocity over the (inner or
outer) surface of the annular (cosmic) string, which is determined
by the boundary condition. We noticed that  a general boundary
condition for transport over an interface [15] was
\begin{equation}
 \delta u=L_s^0 \dot{\gamma}
 (1-\frac{\dot{\gamma}}{\dot{\gamma}_c})^{-1/2},
\end{equation}
where  $\delta u$ is the velocity jump over the interface, $L_s^0$
is a constant slip length, $\dot{\gamma}_c$ is the critical shear
rate at which the slip length diverges. The slip (velocity)
boundary condition above (related to the slip length) is closely
linked to  the mean free path of the particles together with a
geometry-dependent factor (it is the quantum-mechanical scattering
of Bogoliubov quasiparticles which is responsible for the loss of
transverse momentum transfer to the confined surfaces [17]). The
value of $\dot{\gamma}_c$ is a function of the corrugation of
interfacial energy.  \newline With the slip boundary condition
[15], we can derive the velocity fields and transport rates along
the wavy-rough annular tube below using the verified boundary
perturbation technique [15-16] and dimensionless analysis. We
firstly select $L_s^0$ to be the characteristic length scale and
set $r'=r/L_s^0$, $R_1=r_1/L_s^0$, $R_2=r_2/L_s^0$,
$\epsilon'=\epsilon/L_s^0$. After this, for simplicity, we drop
all the primes. It means, now, $r$, $R_1$, $R_2$ and $\epsilon$
become dimensionless ($\phi$ and $\dot{\gamma}$ also follow). The
wavy boundaries are prescribed as $r=R_2+\epsilon \sin(k\theta)$
and $r=R_1+\epsilon \sin(k\theta+\beta)$ and the presumed steady
transport is along the $z$-direction (annulus-axis direction).
\subsection{Boundary Perturbation}
Along the outer boundary (the same treatment below could also be
applied to
the inner boundary), we have
 $\dot{\gamma}=(d u)/(d n)|_{{\mbox{\small on interfaces}}}$.
Here, $n$ means the  normal. Let $u$ be expanded in $\epsilon$ :
 $$u= u_0 +\epsilon u_1 + \epsilon^2 u_2 + \cdots,$$
and on the boundary, we expand $u(r_0+\epsilon dr,
\theta(=\theta_0))$ into
\begin{displaymath}
u(r,\theta) |_{(r_0+\epsilon dr,\,\theta_0)} =u(r_0,\theta)+\epsilon
[dr \,u_r (r_0,\theta)]+ \epsilon^2 [\frac{dr^2}{2}
u_{rr}(r_0,\theta)]+\cdots=
\end{displaymath}
\begin{equation}
  \{u_{slip} +\frac{\dot{\gamma} R_2}{\phi} [\cosh \phi - \cosh (\frac{\phi
 r}{R_2})]\}|_{{\mbox{\small on interfaces}}}, \hspace*{6mm} r_0 \equiv
 R_1, R_2;
\end{equation}
where
\begin{equation}
 u_{slip}|_{{\mbox{\small on interfaces}}}=L_s^0 \{\dot{\gamma}
 [(1-\frac{\dot{\gamma}}{\dot{\gamma}_c})^{-1/2}]\}
 |_{{\mbox{\small on interfaces}}}, 
\end{equation}
Now, on the outer interface (cf. [16])
\begin{displaymath}
 \dot{\gamma}=\frac{du}{dn}=\nabla u \cdot \frac{\nabla (r-R_2-\epsilon
\sin(k\theta))}{| \nabla (r-R_2-\epsilon \sin(k\theta)) |}
=[1+\epsilon^2 \frac{k^2}{r^2}  \cos^2 (k\theta)]^{-\frac{1}{2}}
[u_r |_{(R_2+\epsilon dr,\theta)} -
\end{displaymath}
\begin{displaymath}  
 \hspace*{12mm} \epsilon \frac{k}{r^2}
\cos(k\theta) u_{\theta} |_{(R_2+\epsilon dr,\theta)}
]=u_{0_r}|_{R_2} +\epsilon [u_{1_r}|_{R_2} +u_{0_{rr}}|_{R_2}
\sin(k\theta)-
\end{displaymath}
\begin{displaymath}
  \hspace*{12mm}  \frac{k}{r^2} u_{0_{\theta}}|_{R_2} \cos(k\theta)]+\epsilon^2 [-\frac{1}{2} \frac{k^2}{r^2} \cos^2
(k\theta) u_{0_r}|_{R_2} + u_{2_r}|_{R_2} + u_{1_{rr}}|_{R_2} \sin(k\theta)+ 
\end{displaymath}
\begin{equation}
   \hspace*{12mm} \frac{1}{2} u_{0_{rrr}}|_{R_2} \sin^2 (k\theta) -\frac{k}{r^2}
\cos(k\theta) (u_{1_{\theta}}|_{R_2} + u_{0_{\theta r}}|_{R_2}
\sin(k\theta) )] + O(\epsilon^3 ) .
\end{equation}
Considering $L_s^0 \sim R_1,R_2 \gg \epsilon$ case, we also
presume $\sinh\phi \ll \dot{\gamma}_c/\dot{\gamma_0}$.
With equations (1) and (5), using the definition of
$\dot{\gamma}$, we can derive the velocity field ($u$) up to the
second order :
\begin{displaymath}
u(r,\theta)=-(R_2 \dot{\gamma}_0/\phi) \{\cosh (\phi
r/R_2)-\cosh\phi\, [1+\epsilon^2 \phi^2 \sin^2 (k\theta)/(2
R_2^2)]+
\end{displaymath}
\begin{displaymath}
 \hspace*{12mm} \epsilon \phi \sinh \phi \,
\sin(k\theta)/R_2\}+u_{slip}|_{r=R_2+\epsilon \sin (k\theta)}.
\end{displaymath}
The key point is to firstly obtain the slip velocity along the
boundaries or surfaces.
After lengthy mathematical manipulations, we obtain %
the velocity fields (up to the second order) and then we can
integrate them with respect to the cross-section to get the transport (volume
flow) rate ($Q$, also up to the second order here) :
\begin{displaymath} 
  Q=\int_0^{\theta_p} \int_{R_1+\epsilon \sin(k\theta+\beta)}^{R_2+\epsilon \sin(k\theta)}
 u(r,\theta) r
 dr d\theta =Q_{0} +\epsilon\,Q_{p_0}+\epsilon^2\,Q_{p_2}.
\end{displaymath}
In fact, the approximate (up to the second order) net transport (volume flow)
rate  reads :
\begin{displaymath}
 Q=\pi \dot{\gamma}_0 \{L_s^0 (R_2^2-R_1^2)  \sinh\phi \,
 (1-\frac{\sinh\phi}{\dot{\gamma}_c/\dot{\gamma_0}})^{-1/2}+
 \frac{R_2}{\phi}[(R_2^2-R_1^2)\cosh\phi-\frac{2}{\phi}(R_2^2 \sinh \phi-
\end{displaymath}
\begin{displaymath}
  R_1 R_2 \sinh(\phi \frac{R_1}{R_2}))+ \frac{2 R_2^2}{\phi^2}(\cosh\phi-\cosh(\phi \frac{R_1}{R_2}))]\}+
  \epsilon^2 \{\frac{\pi}{2} u_{slip_0} (R_2^2-R_1^2)+
\end{displaymath}
\begin{displaymath}
 L_s^0 \frac{\pi}{4} \dot{\gamma}_0  \sinh \phi (1+\frac{\sinh\phi}{\dot{\gamma}_c/\dot{\gamma_0}})
 (-k^2+\phi^2)[1-(\frac{R_1}{R_2})^2]+\frac{\pi}{2}\dot{\gamma}_0
 [R_1 \sinh (\frac{R_1}{R_2} \phi)-R_2 \sinh \phi]-
\end{displaymath}
\begin{displaymath}
 \frac{\pi}{2}  \dot{\gamma}_0 \frac{R_2}{\phi} [ \cosh\phi - \cosh (\phi
  \frac{R_1}{R_2})]+ \frac{\pi}{4}  \dot{\gamma}_0 \phi \cosh\phi [R_2  - \frac{R_1^2}{R_2}
  ]+
\end{displaymath}
\begin{displaymath}
  \pi \dot{\gamma}_0 \{[\sinh\phi+L_s^0 \cosh\phi
  (1+\frac{\sinh\phi}{\dot{\gamma}_c/\dot{\gamma_0}})] (R_2-R_1 \cos\beta
  )\}+\frac{\pi}{2}\dot{\gamma}_0 \frac{R_2}{\phi} \cosh \phi+
\end{displaymath}
\begin{equation}
  L_s^0\frac{\pi}{4} \phi^2 \dot{\gamma}_0 \frac{\cosh\phi}{\dot{\gamma}_c/\dot{\gamma}_0}[1
-(\frac{R_1}{R_2})^2
 ]\} \cosh\phi.
\end{equation}
Here,
\begin{equation}
 u_{{slip}_0}= L_s^0 \dot{\gamma}_0 [\sinh\phi(1-\frac{\sinh\phi}{
 \dot{\gamma}_c/\dot{\gamma}_0})^{-1/2}].
\end{equation}
\section{Results and Discussions}
With above detailed derivations,  now, we firstly check the
presumed wavy-roughness effect (or combination of curvature and
confinement effects)  upon the possible frictionless  transport of
many condensed (composite) fermions because there are no available
experimental data and numerical simulations for the same geometric
configuration (annular tube with wavy corrugations in transverse
direction). With a series of forcings (due to externally imposed
gravity or electric field forcings) : $\phi\equiv - R_2 (\delta
{\cal G})/(2\tau_0)$, we can determine the enhanced shear rates
($d\gamma/dt$) due to these forcings. From equation (5), we have
(up to the first order)
\begin{equation}
 \frac{d\gamma}{dt}=\frac{d\gamma_0}{dt} [ \sinh \phi+\epsilon
 \sin(k\theta) \frac{\phi}{R_2} \cosh \phi].
\end{equation}
The parameters
are fixed below (the orientation effect : $\sin(k\theta)$ is fixed
here). $r_2$ (the mean outer radius) is selected as the same as
the slip length $L_s^0$. The amplitude of wavy roughness
can be tuned easily.
The effect of wavy-roughness is significant once the forcing
($\phi$) is rather large (the maximum is of the order of magnitude
of $\epsilon [\phi \tanh(\phi)/R_2]$). \newline If we select a
(fixed) temperature,
 then from the expression of $\tau_0$, we
can obtain the shear stress $\tau$ corresponding to above gravity
forcings ($\phi$) :
\begin{equation}
 \tau =\tau_0 \sinh^{-1} [\sinh(\phi)+\epsilon
 \sin(k\theta) \frac{\phi}{R_2} \cosh(\phi)].
\end{equation}
There is no doubt that the orientation effect ($\theta$) is also
present for the condensed many-fermion system. For illustration
below, we only consider the maximum case : $|\sin(k\theta)|=1$.
The wave number of roughness in transverse direction is fixed to
be a constant.
\newline
As the primary interest of present study is related to the
possible frictionless transport  or formation of superfluidity
(presumed to be relevant to the many-fermion system as mentioned
in Introduction) due to strong shearing, we shall present our main
results in the following. Note that, based on the
absolute-reaction-rate Eyring model (of stress-biased thermal
activation), structural rearrangement is associated with a single
energy barrier (height) $\Delta E$ that is lowered or raised
linearly by a (shear) yield stress $\tau$. If the transition rate
is proportional to the plastic (shear) strain rate (with a
constant ratio : $C_0$; $\dot{\gamma} = C_0 R_t$, $R_t$ is the
transition rate in the direction aided by stress), we have
\begin{equation}
 \tau = 2 [\frac{\Delta E}{V_h} + \frac{k_BT}{V_h} \ln(\frac{\dot{\gamma}}{C_0
 \nu_0})]  \hspace*{28mm} \mbox{if}\hspace*{6mm} \frac{V_h
 \tau}{k_B T} \gg 1
\end{equation}
where $\nu_0$ is an attempt frequency or transition rate, $C_0
\nu_0 \sim  \dot{\gamma}_0 \exp(\Delta E/k_B T)$, or
\begin{equation}
 \tau = 2\frac{k_BT}{V_h} \frac{\dot{\gamma}}{C_0
 \nu_0}\exp({\Delta E}/{k_B T})   \hspace*{24mm} \mbox{if}\hspace*{6mm} \frac{V_h
 \tau}{k_B T} \ll 1.
\end{equation}
It is possible that the frictional resistance (or shear stress)
can be almost zero (existence of $\tau \sim 0$) from above
equations (say, equation (10) considering a sudden jump of the
resistance). The nonlinear character only manifests itself when
the magnitude of the applied stress times the activation volume
becomes comparable or greater in magnitude than the thermal
vibrational energy.
\newline Normally, the value of $V_h$ is associated with a typical
volume required for a microscopic shear rearrangement. Thus, the
nonzero transport rate (of the condensed many-fermion  system) as
forcing is absent could be related to a barrier-overcoming or
tunneling for shear-thinning matter along the wavy-roughness
(geometric valley and peak served as atomic potential surfaces) in
annular tubes when the wavy-roughness is present. Once the
geometry-tuned potentials (energy) overcome this barrier, then the
tunneling (spontaneous transport) inside wavy-rough annular tubes
occurs.
\newline
Finally, We demonstrate in Fig. 1 that if we select the activation
energy to be $4\times 10^{-19}$ J  we can then observe a sudden
drop of the resistance (frictional or shear stress) around 3 order
of magnitude at $T=300.5 ^{\circ}$K ($V_h \approx 3.12 \times
10^{-21}$m$^3$). It means there is a rather low resistance  below
the temperature : $T \sim 300 ^{\circ}$K for the material
parameters selected. As $\tau \sim 0$ (below $T\sim 300$ K), from
$|\tau|= r_2 \delta {\cal G} /2$ ($r_2 \not =0$), we can
understand that there is no  need for any external (gravity or
electric field) forcing ($\delta {\cal G} \sim 0$) once the
persistent current occurs.
\newline
The possible reasoning for this frictionless transport of many
condensed (composite) fermions can be stressed again as a brief
summary. It could be due to the strong shearing driven by larger
external (say, gravity or electric field) forcings along a
confined wavy-rough tube. The shear-thinning (the viscosity
diminishes with increasing shear rate) reduces the viscosity for
the transport of this condensed many-fermion system significantly.
One possible outcome for almost vanishing viscosity is the nearly
frictionless transport. We shall investigate other relevant issues
[18-20] in the future.

\vspace*{20mm}

\newpage

\psfig{file=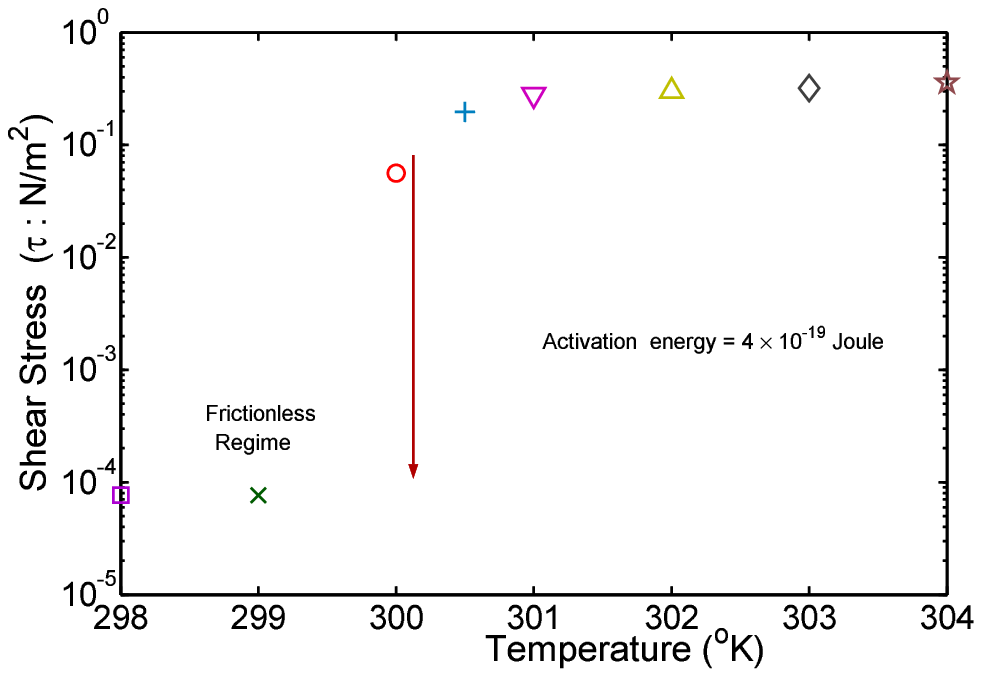,bbllx=-0.5cm,bblly=19cm,bburx=14cm,bbury=27cm,rheight=8cm,rwidth=12cm,clip=}

\begin{figure}[h]
\hspace*{10mm} Fig. 1 \hspace*{1mm} Calculated (shear) stresses or
resistance using an activation energy
 $4 \times 10^{-19}$ J.  \newline \hspace*{8mm} There is a sharp
decrease of shear stress around T $\sim 300.5 ^{\circ}$K. Below
around $300$ K \newline \hspace*{8mm} ($V_h \approx 3.12 \times
10^{-21}$m$^3$), the
 transport of many composite (condensed) fermions is nearly
\newline \hspace*{8mm} frictionless.
\end{figure}
\end{document}